\documentclass[twocolumn,superscriptaddress,aps,noeprint,prb]{revtex4-2}
\usepackage{graphicx}
\usepackage{times}
\usepackage{textcomp}
\usepackage{units}
\usepackage{stmaryrd} 
\usepackage{todonotes}

\usepackage{amssymb}
\usepackage{amsmath}
\usepackage{amsfonts}
\usepackage{physics}
\usepackage{bm}

\usepackage{mathtools}
\usepackage{etoolbox}

\usepackage[nolist,nohyperlinks]{acronym}
\usepackage{hyperref}
\hypersetup{colorlinks=false, allcolors=black, pdfborder={0 0 0}}
\usepackage[noabbrev,capitalize]{cleveref}

\Crefname{equation}{Eq.}{Eqs.}
\Crefname{figure}{Fig.}{Figs.}
\Crefname{tabular}{Tab.}{Tabs.}

\newcommand{\up}{\uparrow}
\newcommand{\dn}{\downarrow}

\newcommand{\sharm}[3][]{\ifblank{#1}{Y^{#2}_{#3}}{\prescript{}{#1}Y^{#2}_{#3}}}

\AtBeginDocument{ 
\def\[#1\]{\begin{align}#1\end{align}}		
\def \<#1\>{\begin{aligned}#1\end{aligned}}
\def\(#1\){\begin{align*}#1\end{align*}} 	
}

\begin{document}

\preprint{0}

\title{Constraints on the two-dimensional pseudo-spin 1/2 Mott insulator
  description of Sr$_2$IrO$_4$}

\author{B. Zwartsenberg}
\affiliation{Quantum Matter Institute, University of British Columbia, Vancouver, BC V6T 1Z4, Canada}
\affiliation{Department of Physics and Astronomy, University of British Columbia, Vancouver, BC V6T 1Z1, Canada}
\author{R. P. Day}
\affiliation{Quantum Matter Institute, University of British Columbia, Vancouver, BC V6T 1Z4, Canada}
\affiliation{Department of Physics and Astronomy, University of British Columbia, Vancouver, BC V6T 1Z1, Canada}
\author{E. Razzoli}
\affiliation{Quantum Matter Institute, University of British Columbia, Vancouver, BC V6T 1Z4, Canada}
\affiliation{Department of Physics and Astronomy, University of British Columbia, Vancouver, BC V6T 1Z1, Canada}
\author{M. Michiardi}
\affiliation{Quantum Matter Institute, University of British Columbia, Vancouver, BC V6T 1Z4, Canada}
\affiliation{Department of Physics and Astronomy, University of British Columbia, Vancouver, BC V6T 1Z1, Canada}
\affiliation{Max Planck Institute for Chemical Physics of Solids, N{\"o}thnitzer Stra{\ss}e 40, 01187 Dresden, Germany}
\author{M.X. Na}
\affiliation{Quantum Matter Institute, University of British Columbia, Vancouver, BC V6T 1Z4, Canada}
\affiliation{Department of Physics and Astronomy, University of British Columbia, Vancouver, BC V6T 1Z1, Canada}
\author{G. Zhang}
\affiliation{Key Laboratory of Materials Physics, Institute of Solid State Physics, HFIPS, Chinese Academy of Sciences, Hefei 230031, People's Republic of China }
\author{J.D. Denlinger}
\affiliation{Advanced Light Source, Lawrence Berkeley National Laboratory,
  Berkeley, California 94720, USA}
\author{I. Vobornik}
\affiliation{Istituto Officina dei Materiali (IOM)-CNR, Laboratorio TASC, in Area Science Park, S.S.14, Km 163.5, I-34149 Trieste, Italy}
\author{C. Bigi}
\affiliation{Dipartimento di Fisica, Universitá di Milano, Via Celoria 16, I-20133 Milano, Italy}
\author{B.J. Kim}
\affiliation{Department of Physics, Pohang University of Science and Technology, Pohang 790-784, South Korea}
\affiliation{Center for Artificial Low Dimensional Electronic Systems, Institute
  for Basic Science (IBS), 77 Cheongam-Ro, Pohang, 790-784, Republic of Korea}
\affiliation{Max Planck Institute for Solid State Research, Heisenbergstra{\ss}e 1, D-70569 Stuttgart, Germany}
\author{I.S. Elfimov}
\affiliation{Quantum Matter Institute, University of British Columbia, Vancouver, BC V6T 1Z4, Canada}
\affiliation{Department of Physics and Astronomy, University of British Columbia, Vancouver, BC V6T 1Z1, Canada}
\author{E. Pavarini}
\email{e.pavarini@fz-juelich.de}
\affiliation{
Institute for Advanced Simulation, Forschungszentrum J\"ulich,
D-52425 J\"ulich, Germany}
\affiliation{JARA High-Performance Computing, Forschungszentrum J\"ulich, J\"ulich, Germany}
\author{A. Damascelli}
\email{damascelli@physics.ubc.ca}
\affiliation{Quantum Matter Institute, University of British Columbia, Vancouver, BC V6T 1Z4, Canada}
\affiliation{Department of Physics and Astronomy, University of British Columbia, Vancouver, BC V6T 1Z1, Canada}

\begin{abstract}
  Sr$_{2}$IrO$_{4}$ has often been described via a simple, one-band pseudo-spin
  1/2 model, subject to electron-electron interactions, on a square lattice,
  fostering analogies with cuprate superconductors, believed to be well
  described by a similar model. In this work we argue -- based on a detailed
  study of the low-energy electronic structure by circularly polarized spin and
  angle-resolved photoemission spectroscopy combined with dynamical mean-field
  theory calculations -- that a pseudo-spin 1/2 model fails to capture the full
  complexity of the system. We show instead that a realistic multi-band Hubbard
  Hamiltonian, accounting for the full correlated $t_{2g}$ manifold, provides a
  detailed description of the interplay between spin-orbital entanglement and
  electron-electron interactions, and yields quantitative agreement with
  experiments. Our analysis establishes that the $j_{3/2}$ states make up a
  substantial percentage of the low energy spectral weight, i.e.\ approximately
  74\% as determined from the integration of the $j$-resolved spectral function
  in the $0$ to $-1.64$ eV energy range. The results in our work are not only of
  relevance to iridium based materials, but more generally to the study of
  multi-orbital materials with closely spaced energy scales.
  \\
\end{abstract}

\maketitle

\begin{acronym}[ARPES]
\acro{ARPES}[ARPES]{angle-resolved photoelectron spectroscopy}
\acro{BZ}{Brillouin zone}
\acro{MDC}{momentum distribution curve}
\acro{EDC}{energy distribution curve}
\acro{RIXS}{resonant inelastic x-ray scattering}
\acro{SOC}{spin-orbit coupling}
\acro{STM}{scanning tunnelling microscopy}
\acro{STS}[STS]{scanning tunnelling microscopy}
\acro{BCT}{body centred tetragonal}
\acro{DMFT}{dynamical mean field theory}
\end{acronym}

\section{Introduction}

Sr$_{2}$IrO$_{4}$ has been studied since shortly after the discovery of the
cuprate superconductors \cite{Bednorz1986, Cava1994}, as the compound was
believed to share some of its defining properties with the copper oxides. More
specifically, Sr$_{2}$IrO$_{4}$ shares its structure with the superconducting
``parent compound'' La$_2$CuO$_4$, and it features a similar anti ferromagnetic
ground state \cite{Cava1994, Crawford1994, Cao1998}. A key difference is that
the cuprates are described by a single hole in the $e_{g}$ manifold, as opposed
to the iridate that has a single hole in the $t_{2g}$ manifold. In the seminal
work by Kim \emph{et al.}, it was suggested that the $t_{2g}$ orbitals entangle
into a filled $j_{\mathrm{eff}} = 3/2$, and a half filled
$j_{\mathrm{eff}} = 1/2$ manifold \cite{Kim2008}. It was quickly realized that
this scenario would bring Sr$_2$IrO$_4$ even closer to the quintessential
cuprate superconductor: a (pseudo-) spin 1/2 Mott insulator on a square
two-dimensional lattice. Theoretical calculations predicted a superconducting
state may exist in a $j_{\mathrm{eff}}$ pseudo-spin 1/2 system when electron
doped \cite{Wang2011}, with more sophisticated analyses including all $t_{2g}$
orbitals and strong spin-orbit coupling painting a similar picture
\cite{Watanabe2013, Meng2014}.
Promising observations were made in experiments: it was found that the
excitations of the pseudospins probed by \ac{RIXS} are reminiscent of a
Heisenberg model \cite{Kim2012a,Kim2012b}, the expected low energy behaviour for
a spin 1/2 Mott insulator \cite{Kastner1998, Birgeneau2006}. In addition,
features reminiscent of doped Mott insulators, such as a v-shaped gap and a
phase separated spatial distribution, were seen in \ac{STM} \cite{Battisti2016},
and a pseudogap was detected in angle-resolved photoemission spectroscopy
(ARPES) \cite{DeLaTorre2015}. Even stronger evidence was found in surface doped
samples: \ac{STM} and ARPES observe a gap that reminds of those found in
cuprate superconductors \cite{Yan2015, Kim2015}. However, these are
spectroscopic observations that are constrained to the surface, and so far no
signatures of bulk superconducting behaviour have been reported in the literature. \\

\begin{figure*}
\centering
  \includegraphics[width = 7in]{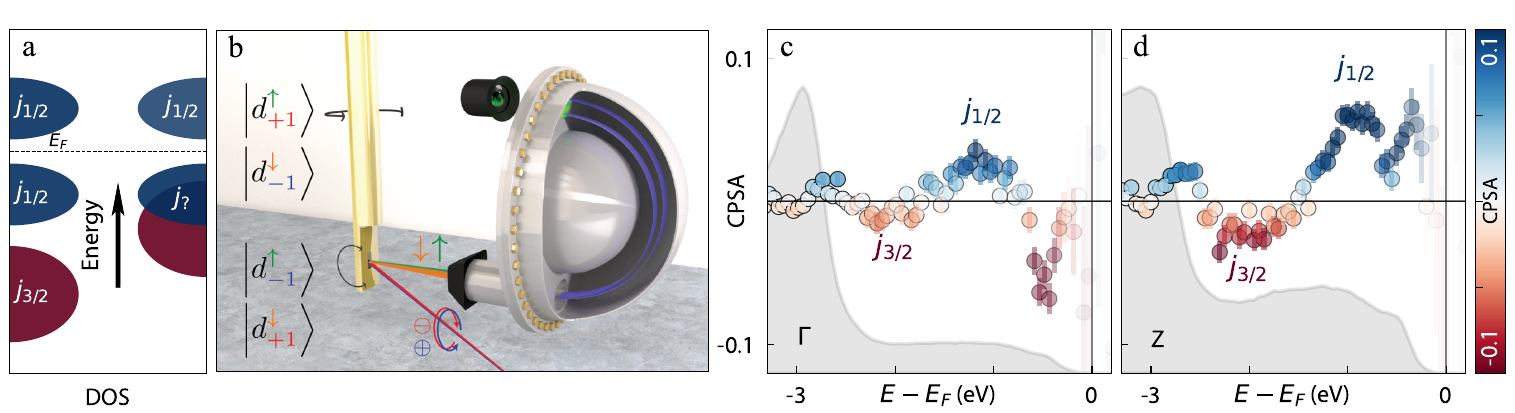}
  \caption{ (a) Schematic representation of a fully decoupled
    pseudo-spin 1/2 model (left), and an entangled multi-orbital Mott system
    (right). (b) Schematic depiction of the circularly polarized spin-ARPES
    (CPS ARPES) experiment. Different spin-orbitally entangled states
    $\ket{d_{m_\ell}^{m_s}}$ can be selected by choosing a combination of
    circular polarization $\{\oplus,\ominus\}$ and spin-detector channel
    $\{\uparrow,\downarrow\}$. (c,d) The CPS ARPES intensity obtained at normal emission using 51.1 (c) and 64 (d) eV photons
    corresponding to the $\Gamma$ and $\mathrm{Z}$ points in the Brillouin zone respectively (colored markers), and the sum of spin and
    polarization dependent signals (grey shaded).}\label{fig:CPSA_higheb}
\end{figure*}
A potential factor in the explanation for the lack of superconductivity may be
found in the non-trivial departure from a simple spin 1/2 scenario. We start by
pointing out that the theoretical models predicting superconductivity have been
derived in the strong spin-orbit coupling limit \cite{Wang2011,Watanabe2013,
  Meng2014} (i.e. in the limit of a ``simple'' pseudo-spin 1/2 model). Although
spin-orbit coupling is large in this system ($\sim 0.45$ eV
\cite{Mattheiss1976,Moon2008,Kim2012b}), it is still modest compared to the
overall bandwidth ($\sim 2$ eV) of the $t_{2g}$ bands
\cite{Brouet2015,Cao2016,Louat2018}. A complete splitting into $j_{3/2}$ and
$j_{1/2}$ \footnote{To reduce the notational clutter, we will henceforth refer
  to the $j_{\mathrm{eff}} = 1/2$ and $3/2$ states as $j_{1/2}$ and $j_{3/2}$
  respectively} multiplets is therefore likely not realized, and there has been
some sporadic evidence that supports this idea. It was pointed out that toward
the Brillouin zone boundaries, the pristine $t_{2g}$ character dominates the
spin-orbital entanglement, and not much mixing occurs~\cite{Louat2018}. Neutron
scattering shows that the local moments are far from the idealized $j_{1/2}$
picture, and in reality the eigenstates bear more resemblance to a $d_{xy}$
orbital \cite{Jeong2020}. Taken altogether, these arguments suggest that a
pseudo-spin 1/2 model may not be a sufficient description of the system, and
raise questions about the true nature of the
Sr$_{2}$IrO$_{4}$ ground state. \\

In this paper, we use a technique that can directly attend to the question of
whether a pseudo-spin 1/2 model is indeed a valid description for
Sr$_{2}$IrO$_{4}$. In order to do this, we measure the spin-orbital entanglement
of the valence band states, i.e. the expectation value
$\expval{\vb{L}\cdot\vb{S}}$ for the low energy states, using circularly
polarized spin-ARPES (CPS-ARPES) and observe a clear departure from the
canonical $j_{1/2}$ model. We are instead able to explain our observations using
\ac{DMFT} calculations, which accurately predict the non-trivial behavior of
this multi-band, spin-orbit coupled Mott system, without imposing a predefined
hierarchy onto the magnitudes of these effects. Our conclusion is that a spin
1/2 model is insufficient to capture all intricacies of the low energy
electronic structure of this material, but more importantly that we gain a
strong understanding of Sr$_{2}$IrO$_{4}$ by
comparing our experiments to an adequately powerful theoretical description.\\

To make substantiated arguments about the sufficiency of the $j_{1/2}$ model,
the quantum number $j$ should be measured for the low energy manifold, giving a
distinct character for the $j_{1/2}$ and $j_{3/2}$ states. If the system can be
described as a pseudo-spin 1/2 system, the $j_{3/2}$ states must be far enough
into the valence bands so that they do not overlap with, or couple to, the
$j_{1/2}$ states [Fig. \ref{fig:CPSA_higheb} (a), left]. A sizeable overlap or
coupling would result in bands with both $j_{1/2}$ and
$j_{3/2}$ character [Fig. \ref{fig:CPSA_higheb} (a), right]. However, while the
quantum number $j$ is not directly accessible in ARPES measurements, the alignment
of spin and orbital angular momentum $\expval{\vb{L}\cdot\vb{S}}$, which has an
immediate relation to $j$, can be measured directly. For a
pure $j_{1/2}$ state this quantity should be positive
($\expval{\vb{L}\cdot\vb{S}} = 1$), while it would be negative for a pure
$j_{3/2}$ state ($\expval{\vb{L}\cdot\vb{S}} = -\frac{1}{2}$) \footnote{Since
  the $j_{\mathrm{eff}}$ states are constructed from $\ell_{\mathrm{eff}} = -1$,
  the 1/2 (3/2) state has parallel (antiparallel) alignment of spin and orbital
  angular momentum. A more thorough review is presented in the appendix.}. \\

\section{Circularly Polarized Spin ARPES}
To quantify the spin-orbital entanglement, spin-resolved measurements are
performed using circularly polarized light, as schematically depicted in
\cref{fig:CPSA_higheb} (b). This technique has been used previously in
angle-integrated photoemission \cite{Pierce1976, Mizokawa2001}, as well as in
ARPES on Sr$_{2}$RuO$_{4}$ \cite{Veenstra2014} and iron pnictides
\cite{Day2018}. The use of circularly polarized light selects a particular
$m_\ell$ value $\{-1, +1\}$, through the photoemission dipole matrix element,
while the spin-detector selects between states with
$m_s = \{\uparrow,\downarrow\}$. By combining these two filters, and measuring
the four individual components, it is possible to obtain the spin-orbital
entanglement. In particular, it can be shown that at normal emission, the $z$
component of $\expval{\vb{L}\cdot\vb{S}}$, i.e. $\expval{L_z S_z}$, can be recovered.
\begin{figure*}[t]
\centering
  \includegraphics[width = 7in]{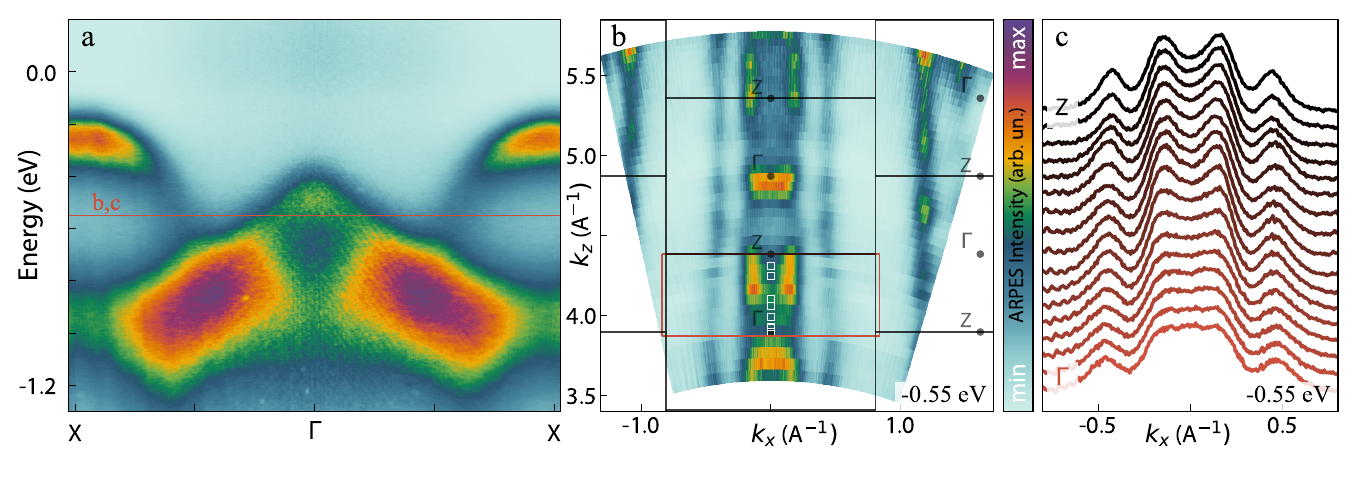}
  \caption{ Overview of photon energy dependent results in Sr$_2$IrO$_4$. (a)
    Spectrum at $h\nu = 100$ eV ($k_{z} = 5.29$ \AA$^{-1}$). The red line
    indicates the energy ($E = -0.55$ eV) at which the plots presented in panels (b) and (c) have
    been generated. (b) Constant energy map at $E = -0.55$ eV, using a sum of
    $\sigma$ and $\pi$-polarized light. The white boxes indicate the positions
    of later presented CPS ARPES data. (c) A series of MDCs measured with photon
    energies ranging from $h\nu = 51$ eV ($k_{z} = 3.89$ \AA$^{-1}$) to
    $h\nu = 84$ eV ($k_{z} = 4.87$ \AA$^{-1}$), with $\sigma$-polarization at
    $E = -0.55$ eV; the ($k_{x},k_{z}$) area encompassed is indicated by a red box in (b). }
\label{fig:CE_and_MDCs}
\end{figure*}

To derive this property, we start by considering the photoemission dipole matrix
element, arising from Fermi's golden rule (for a thorough review, the reader is
referred to
\cite{Day2019})
\begin{multline}
\label{eqn:matrixelement}
  M_{i,f}^{\varepsilon,\sigma,k}  =
  \mel{\psi_{f}^k}{\vb{r}\cdot\boldsymbol{\varepsilon}}{\phi_{i}} =\\
  \sum_{\substack{\ell_i,\ell_f,m_i\\m_\varepsilon,m_f} }
  c_{l_{i}}^{m_{i}} B_{n_i,\ell_i,\ell_f}
  \mel{\sharm{m_i}{\ell_i}}
  {\sharm{m_\varepsilon}{1}}{\sharm{m_f}{\ell_f}}
  \sharm{m_f}{\ell_f} \left( \theta_\mathbf{k}, \phi_\mathbf{k} \right).
\end{multline}
  with $\ket{\psi_{f}^{k}}$ and $\ket{\phi_{i}}$ the final and initial states
 respectively, $\boldsymbol{\varepsilon}$ the polarization vector,
 $c_{l_{i}}^{m_{i}}$ the initial state coefficient in the basis of spherical
 harmonics, and $B_{n_{i}, \ell_{i}, \ell_{f}}$ a radial integral:
 \[
   \label{eqn:bnell_form}
B_{n_{i}, \ell_{i}, \ell_{f}} = \int dr r^3 R_{n_i,\ell_i}(r) j_{\ell_f}(r),
\]
where $R_{n_{i}, \ell_{i}}(r)$ is the radial part of the basis functions and
$j_{\ell_{f}}(r)$ are the spherical Bessel functions. Using circularly polarized
light with positive helicity gives
$\boldsymbol{\varepsilon}^{\oplus}\cdot\vb{r} = \varepsilon_0 \left(x + i y\right) = \varepsilon_0 \sharm{1}{1}$.
The matrix element then becomes:
\begin{multline}
\label{eqn:cpsa_melem}
M_{i,f}^k  = \mel{\psi_{f}^k}{\vb{r}\cdot\boldsymbol{\varepsilon}}{\phi_{i}} = \\
\varepsilon_0  \sum_{\substack{\ell_f \\ m_f, m_i}}
c_{l_{i}}^{m_{i}}B_{n_i,\ell_i,\ell_f}
\mel{\sharm{m_f}{\ell_f}}{\sharm{1}{1}}{\sharm{m_i}{\ell_i}}
\sharm{m_f}{l_f}\left(\theta_k, \phi_k\right),
\end{multline}
At the $\Gamma$ point, we can simplify this equation by using the fact that the
spherical harmonic $\sharm{m_f}{l_f}\left(\theta_k, \phi_k\right)$ has nodes
for all $m_f$ except $m_f = 0$, where its value is 1. With the spherical
harmonic arising from the polarization vector set to $\sharm{1}{1}$, we only
emit from a single initial state spherical harmonic. We can therefore simplify
the expression in \cref{eqn:cpsa_melem} to:
\begin{multline}\label{eqn:asdf}
  M_{i,f}^{k\sigma}  =
  \varepsilon_0  \sum_{\ell_f}
  c^{m_i = -1,\sigma}_{\ell_i}B_{n_i,\ell_i,\ell_f}
  \mel{\sharm{0}{\ell_f}}{\sharm{1}{1}}{ \sharm{-1}{\ell_i}} .
\end{multline}
Noting that the product of spherical harmonics does not depend on $m_{i}$, we
can take the sum over $\ell_{f}$ up into a constant prefactor. We denote
$a_{\ell_{i},\ell_{f}} = \mel{\sharm{0}{\ell_f}}{\sharm{1}{1}}{ \sharm{-1}{\ell_i}}$. To
get the photoemission intensity, we take the squared norm:
\begin{multline}\label{eqn:bnl_sum}
  I^{\oplus\sigma}  = \varepsilon_0^2 \left( \sum_{\ell_f} B_{n_i,\ell_i,\ell_f} a_{\ell_i,\ell_f} \right)^2 \abs{c^{m_i = -1,\sigma}_{\ell_i}}^2 \\
  = A \abs{c^{-1,\sigma}_{\ell_i}}^2.
\end{multline}
It follows trivially that we can measure the other components using
$\sigma = \up, \dn$ and $\varepsilon = \oplus, \ominus$ to construct:
\begin{multline}\label{eqn:LzSzfromcpsa}
  I^{\ominus\up} - I^{\oplus\up} - I^{\ominus\dn} + I^{\oplus\dn}  \\
  = A \left(\abs{c^{1,\up}}^2  - \abs{c^{-1,\up}}^2 -  \abs{c^{1,\dn}}^2  + \abs{c^{-1,\dn}}^2 \right)
\end{multline}
Noting that in the basis of
$\ket{m_l = 1, \up}$, $\ket{-1, \up}$, $\ket{1, \dn}$, $\ket{-1, \dn}$, we have:
\begin{equation}
L_zS_z =
\frac{\hbar^2}{2}\begin{pmatrix}
                   1  & 0 &  0 & 0 \\
                   0  & -1  & 0  & 0\\
                    0  & 0 &  -1 & 0 \\
                    0  & 0 & 0  & 1\\

\end{pmatrix},
\end{equation}
we get for $\expval{L_zS_z}$:
\begin{equation}
\expval{L_zS_z} = \frac{\hbar^2}{2}\left(\abs{c^{1,\up}}^2  - \abs{c^{-1,\up}}^2 -  \abs{c^{1,\dn}}^2  + \abs{c^{-1,\dn}}^2 \right),
\end{equation}
which is precisely the expression found in \cref{eqn:LzSzfromcpsa}, aside
from the prefactor. Note that the expression derived above is independent (up to
the prefactor $A$) of the values for $B_{n_i,\ell_i,\ell_f}$. Since there is
only a single term of $m_{l_i}$ for each configuration, there are no
interference terms and the sum in \cref{eqn:bnl_sum} can be evaluated
separately.
\vspace{2mm}
This formulation of $\expval{L_zS_z}$ in terms of $I^{\varepsilon,\sigma}$ is
unfortunately only valid if all factors $B_{n_i,\ell_i,\ell_f}$ are the
identical for both polarizations $\varepsilon^{\oplus}$ and
$\varepsilon^{\ominus}$, which may not be the case in a system where there is
circular dichroism. Moreover, if the sensitivity of the spin-detectors is not
equal for up and down channels, the description also breaks down. By denoting
the sensitivity of the detector of each spin detector as $\eta^{\sigma}$, and
the factor related to the circular dichroism as $\alpha^{\varepsilon}$, we can
write the measured photoemission signal as:
\begin{equation}
\tilde{I}^{\varepsilon\sigma} = \alpha^{\varepsilon}\eta^{\sigma} I^{\varepsilon,\sigma} = \alpha^{\varepsilon}\eta^{\sigma} A \abs{c^{m_i,\sigma}_{\ell_i}}^2,
\end{equation}
where $m_i = -1$ for $\varepsilon^{\oplus}$ and 1 for $\varepsilon^{\ominus}$.
Substituting the $\tilde{I}$ into \cref{eqn:LzSzfromcpsa}, the expectation
value $\expval{L_zS_z}$ is no longer recovered as a result of the prefactors. We
can instead take advantage of the geometric mean $P$ which divides out the
prefactors:
%
\begin{figure}[t]
\centering
  \includegraphics[width = 2.5in]{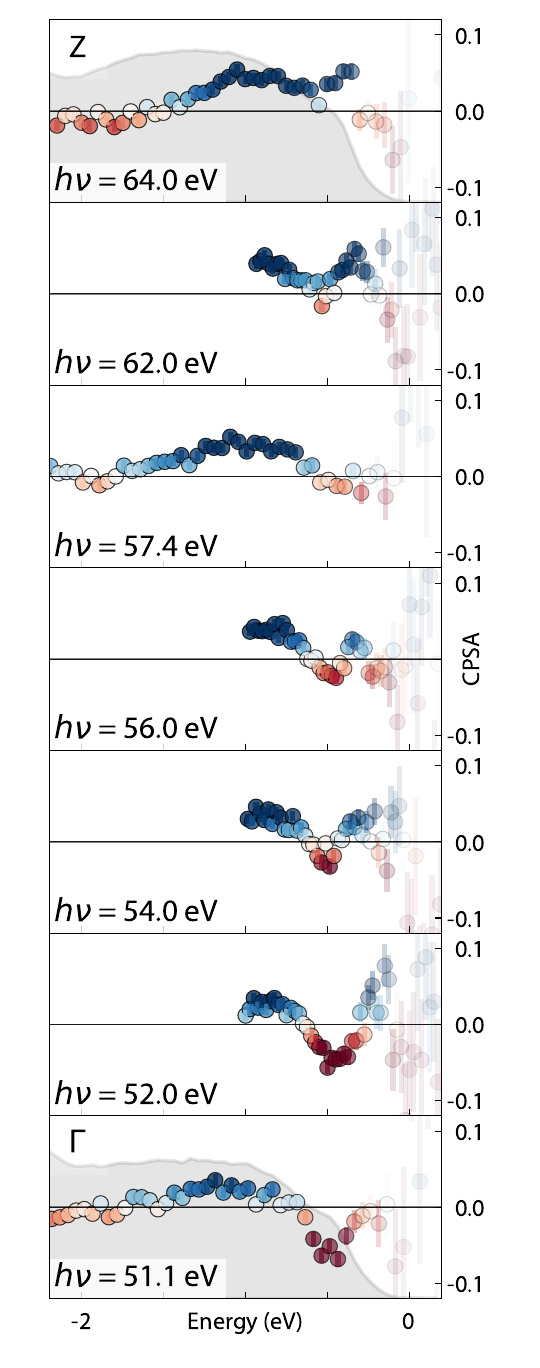}
  \caption{CPS ARPES measurements (colored markers), for photon energies ranging
    from 51.1 eV to 64 eV. The sum of spin and polarization dependent signals
    (grey shaded) corresponds to spin-integrated ARPES.}
\label{fig:CPSA_kz}
\end{figure}
\begin{multline}\label{eqn:geometricmean}
  P = \frac{\sqrt{\tilde{I}^{\ominus\up} \tilde{I}^{\oplus\dn}} - \sqrt{\tilde{I}^{\oplus\up}\tilde{I}^{\ominus\dn}}}{\sqrt{\tilde{I}^{\ominus\up} \tilde{I}^{\oplus\dn}} + \sqrt{\tilde{I}^{\oplus\up}\tilde{I}^{\ominus\dn}}} \\
  =
\frac{\sqrt{\abs{c^{1,\up}}^2\abs{c^{-1,\dn}}^2} - \sqrt{\abs{c^{-1,\up}}^2\abs{c^{1,\dn}}^2}}{\sqrt{\abs{c^{1,\up}}^2\abs{c^{-1,\dn}}^2} + \sqrt{\abs{c^{-1,\up}}^2\abs{c^{1,\dn}}^2}}.
\end{multline}
In the case of Kramers degeneracy, we should have $\abs{c^{m,\sigma}}^2 = \abs{c^{-m,\bar{\sigma}}}^2$, and using the fact that the states are normalized ($\sum \abs{c^{m,\sigma}}^2 = 1$) we obtain:
\begin{multline}\label{eqn:geometricmean_to_P}
P = \frac{\abs{c^{1,\up}}^2 - \abs{c^{1,\dn}}^2}{\abs{c^{1,\up}}^2 + \abs{c^{1,\dn}}^2} \\ = \abs{c^{1,\up}}^2 - \abs{c^{1,\dn}}^2 - \abs{c^{-1,\up}}^2 + \abs{c^{-1,\dn}}^2 = \frac{2}{\hbar^2}\expval{L_zS_z}.
\end{multline}
Using the geometric mean, we can thus extract the expectation value for
$\expval{L_zS_z}$ without the need to know the exact detector sensitivities or
circular dichroism effects. \\

\section{Experimental Results}
Spin-resolved measurements were performed at the VESPA endstation
\cite{Bigi2017} at the Elettra Sincrotrone Trieste, using VLEED spin detectors.
We present the result of applying CPS-ARPES to Sr$_2$IrO$_4$ in
\cref{fig:CPSA_higheb} (c,d), which display the observed CPS ARPES intensity
(colored markers) at normal emission using 51.1 eV ($\Gamma$) and 64 eV ($Z$)
photons respectively. The grey shaded curves represent the sums of all signals
(corresponding to spin-integrated ARPES). Comparing panels (b) and (c,d) we can
readily identify various features: negative ($-2$ eV) and positive ($-1$ eV)
regions, belonging respectively to states with $j_{3/2}$ and $j_{1/2}$
character. Although the data from the $Z$ point in the Brillouin zone
[\cref{fig:CPSA_higheb} (d)] are in line with a simple pseudo-spin 1/2 picture,
the strong negative signal around $E = -0.5$ eV at the $\Gamma$ point
[\cref{fig:CPSA_higheb} (c)] appears to be inconsistent. In the remainder of the
paper we will show that this indeed constitutes a violation of the pseudo-spin 1/2 picture.\\

First we will provide a more detailed analysis along different crystal momenta,
to capture a more complete picture of the spin-orbital entanglement. We note
that while it is possible to measure CPS ARPES along the in-plane momentum
($k_{x}, k_{y}$), data taken this way are much more challenging to interpret. We
have nevertheless measured in-plane CPS ARPES, for which the data and
corresponding analysis can be seen in full in the appendix. Here instead we focus our attention on $k_{z}$, also in
light of the puzzling results in \cref{fig:CPSA_higheb}. In ARPES measurements,
the perpendicular momentum ($k_{z}$) is accessible through changing the incident
photon energy. Although Sr$_2$IrO$_4$ is quasi-two-dimensional, the extended Ir
$5d$ orbitals have the potential to magnify the out of plane hopping. The $k_z$
dispersion in Sr$_2$IrO$_4$ and the related bilayer Sr$_{2}$Ir$_{2}$O$_{7}$
compound has been studied previously \cite{Wang2013}, and a modest energy
dispersion was observed at the $X$ point. However, no data has been presented at
normal emission, which is where our study is concerned. \\

To provide some context for the forthcoming CPS ARPES data, we first consider
spin-integrated photon energy dependent ARPES data. Photon energy dependent
spin-integrated ARPES measurements presented here were taken at the MERLIN
beamline of the Advanced Light Source. Data were acquired between 50 and 120 eV.
The data are corrected using an inner potential~\cite{Damascelli2004} $V_0 = 11$
eV, in good agreement with earlier published results \cite{Wang2013}. We plot a
valence band mapping of Sr$_2$IrO$_4$ along the $\Gamma-\mathrm{X}$ in
\cref{fig:CE_and_MDCs}(a). A constant energy map at $E = -0.55$ eV in the
$k_{z}-k_{x}$ plane is displayed in \cref{fig:CE_and_MDCs}(b), for a sum of
$\pi$- and $\sigma$-polarization. The modulated intensity changes, especially
those periodic in $k_z$, are a clear sign of interlayer coupling, and of an
underlying $k_{z}$ dispersion. A closer inspection reveals pinching of the
cylindrical state around $\Gamma$, which becomes particularly clear when
considering momentum distribution curves (MDCs) between $\Gamma$ and $Z$
[\cref{fig:CE_and_MDCs}(c)].
Although we find clear evidence of $k_z$ dispersion in the exposition of MDCs,
the broad nature of the bands makes observing a simple periodic oscillation in
the corresponding energy distribution curves (EDCs) more challenging.
We also
note that the energy scale appears significantly smaller than the in-plane
bandwidth ($\sim$ 1-2 eV), or even the spin-orbit coupling parameter ($\sim 0.45$ eV);
however, as pointed out in previous work on Sr$_{2}$RuO$_{4}$ \cite{Veenstra2014}, states close to
degeneracies can undergo significant changes as a result of spin-orbit coupling
effects, which in turn can lead to a remarkable $k_{z}$ dependence of the character
of the eigenstates even though a sizable energy dispersion is notably absent. \\

\begin{figure}[t]
 \centering
\includegraphics[width=85mm]{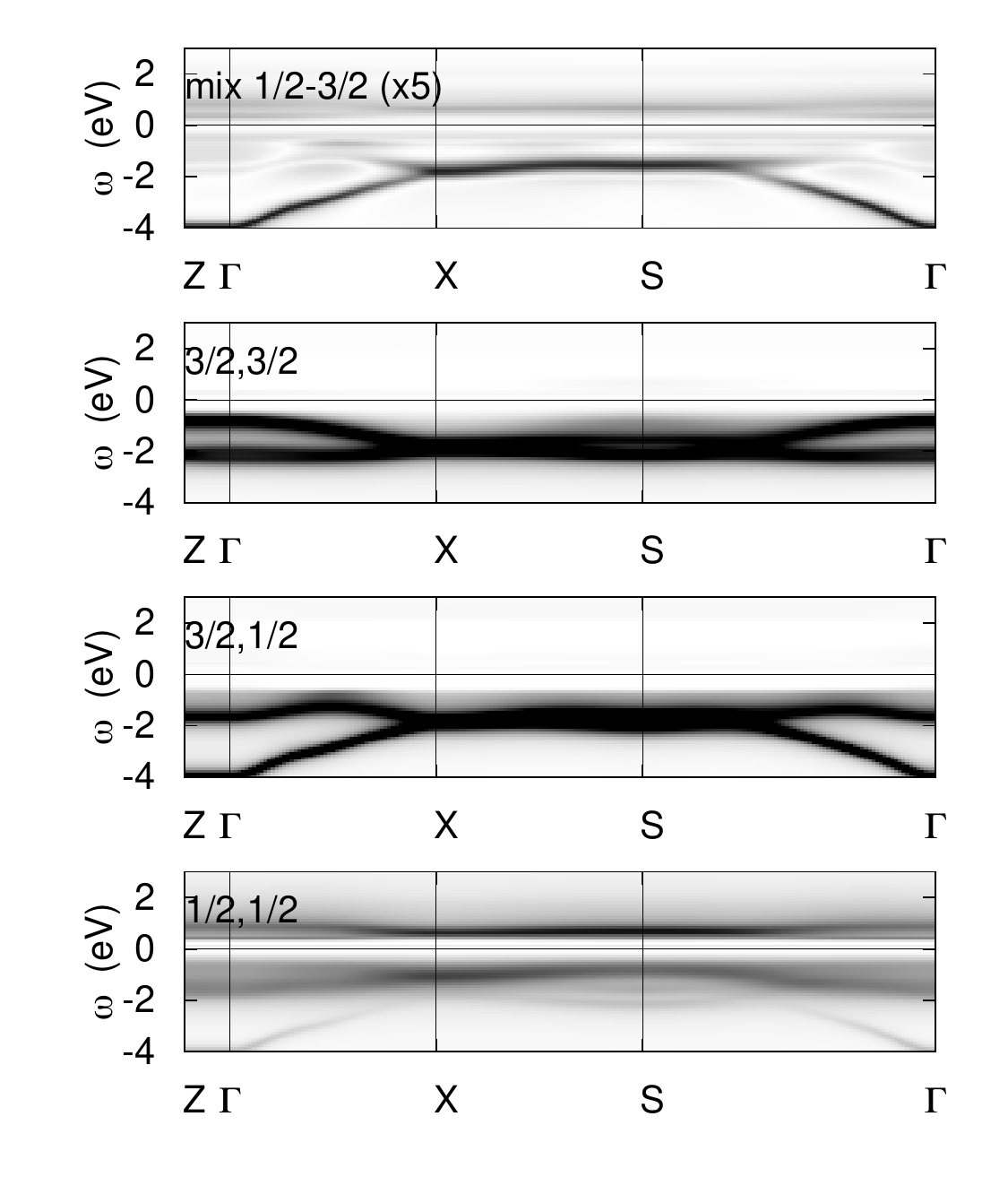}
\caption{\label{fig:dmft_spectrum}  Diagonal and off-diagonal components of the spectral function in the $j$ basis.}
\end{figure}
We now return to CPS ARPES experiments, focusing purely on normal emission
measurements, and changing $k_z$ only through adjusting the photon energy of the
incident beam. We would like to reiterate that in this case the CPS ARPES measurement
is directly proportional to the expectation value of $\expval{L_zS_z}$, and
photoemission matrix element effects cancel out completely.
Photon energy dependent CPS ARPES results are presented in \cref{fig:CPSA_kz} as
colored markers. A grey background indicates the sum of all four individual
spin- and light-polarization dependent signals, which corresponds to
spin-integrated photoemission. The progression of the CPS ARPES signal is
evident, and provides context and additional proof for the puzzling result first
presented in \cref{fig:CPSA_higheb}. Although the positive and negative signal
around $E = -1$ and $-2$ eV is present in all the spectra, the data at low
binding energies paint a contrasting picture. The peak in the spectrum at
$E = -0.5$ eV that starts out negative in \cref{fig:CPSA_kz} at 51.1 eV
($\Gamma$) can be seen to change sign as the photon energy increases to 64 eV
($Z$). It should be stressed that this is an important result: the character of
the spin-orbital entanglement of the lowest-energy band changes from parallel to
antiparallel upon varying $k_z$, revealing a drastic change in the character of
the lowest-energy eigenstates. Neither this sign reversal nor the negative
signal observed at $\Gamma$ are reconcilable with a simple pseudo-spin model,
and require us to
rethink our description of the low-energy states of  Sr$_{2}$IrO$_{4}$. \\

\section{Comparison to DMFT}
We now attempt to shed light on our observations by constructing and solving a model that
goes beyond the pseudo-spin 1/2 framework. To this end, we turn to dynamical mean field theory (DMFT) calculations for
a realistic multi-band Hubbard Hamiltonian.
The method adopted can be summarized as follows. We calculate the electronic
structure (including spin-orbit effects) in the local-density approximation
(LDA) via the full-potential linearized augmented plane-wave method, as
implemented in the WIEN2k code \cite{Blaha2018}. A set of $t_{2g}$ Wannier
functions centered at the Ir atoms and spanning the $t_{2g}$ bands is then
constructed. In this basis we build the system-specific $t_{2g}$ Hubbard model:
\begin{figure}[t]
 \centering
  \includegraphics[width=75mm]{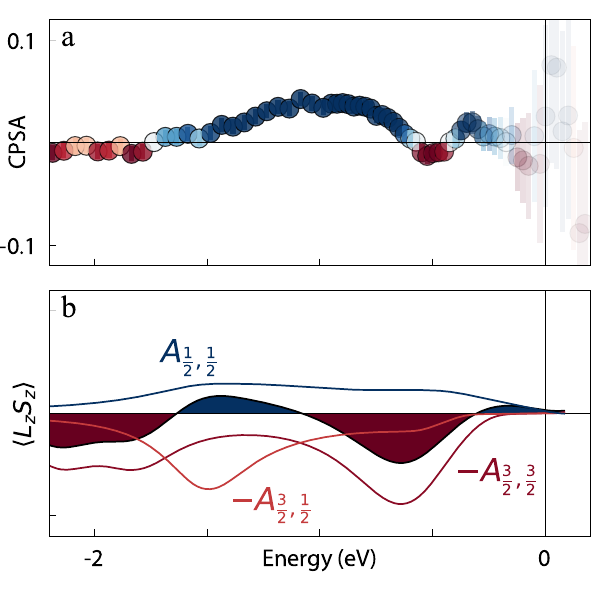}
  \caption{\label{fig:dmftspectral} Comparison between experiment and DMFT. (a)
    $k_{z}$ integrated CPS ARPES data at normal emission. (b) LDA+DMFT
    calculations for Sr$_{2}$IrO$_{4}$ with Coulomb parameters $U = 3.2$ eV and
    $J = 0.4$ eV. Bare lines correspond to diagonal elements
    $A_{j,m_j}({\bf k};\omega)$ of the spin-orbital resolved spectral function
    in the $j$ basis at the $\Gamma$ point. The filled colored curve shows the
    calculated $\expval{L_z S_z}$.}
\end{figure}
\begin{align} \label{Ham} \hat{H}=&-\sum_{ii'}\sum_{mm'}\sum_{\sigma\sigma'}t_{m\sigma,m'\sigma'}^{i,i'}\hat{c}_{im\sigma}^{\dagger}\hat{c}_{i'm'\sigma'} \\ \nonumber &+\frac{1}{2}\sum_i \sum_{mm'pp'}\sum_{\sigma\sigma'}U_{mm'pp'}\hat{c}_{im\sigma}^{\dagger}\hat{c}_{im'\sigma'}^{\dagger}\hat{c}_{ip'\sigma'}\hat{c}_{ip\sigma}.
\end{align} In the Hamiltonian above, $\hat{c}_{im\sigma}^{\dagger}$
($\hat{c}_{im\sigma}$) creates (annihilates) an electron at lattice site $i$
with spin $\sigma \in\{\uparrow,\downarrow \}$ and orbital $m\in\{xy,yz,xz\}$.
The parameters $-t_{m\sigma,m'\sigma'}^{i,i}$ define the on-site crystal-field
matrix, including local spin-orbit terms; the intersite ($i\ne i'$) terms
$-t_{m\sigma,m'\sigma'}^{i,i'}$ are the hopping integrals, also with spin-orbit
interaction contributions. The key screened Coulomb integrals are the direct
Coulomb interaction, $U_{mm'mm'} = U_{m,m'} = U-2J(1-\delta_{m,m'})$, the
exchange Coulomb interaction $U_{mm'm'm} = J$, the pair-hopping term,
$U_{mmm'm'} = J$, and the spin-flip term $U_{mm'm'm} = J$. We adopt the values
$(U,J)=(3.2,0.4)$~eV, corresponding to an average Coulomb repulsion of
$U_{\rm avg}=2.4$~eV; this reproduces the small insulating gap well, as we have
shown in Ref.~\cite{Zhang2021}. We solve (\ref{Ham}) with DMFT using the
interaction-expansion continuous time quantum Monte Carlo impurity solver, in
the implementation developed in
Refs.~\onlinecite{Gorelov2010,Zhang2016,Zhang2017}. The calculations presented
have been performed at the electronic temperature 290K. We obtain the orbital
and ${\bf k}$-resolved spectral-function matrix using the maximum-entropy
method. In \cref{fig:dmft_spectrum} we show the weight of each component along
high-symmetry lines if the Brillouin Zone. The $k_z$ dispersion itself is small
and difficult to resolve, but the shift in character with energy is very clear
at any k point. From here we calculate
\begin{align} {\cal A}_{ \langle L_z S_z \rangle }&= \frac{1}{4} \bigg(A_{1,1}^\uparrow+A_{-1,-1}^\downarrow-A_{1,1}^\downarrow-A_{-1,-1}^\uparrow\bigg)
\end{align} where $A_{m,m}^\sigma$ is the spectral function for orbital $m$ and
spin $\sigma$. In the basis of the $j=1/2$ and $j=3/2$ states
\begin{align} {\cal A}_{ \langle L_z S_z \rangle }&= -\frac{1}{2} A_{\frac{3}{2};\frac{3}{2}}+\frac{1}{6} A_{\frac{3}{2};\frac{1}{2}}+\frac{1}{3} A_{\frac{1}{2};\frac{1}{2}}+\frac{\sqrt{2}}{{3}} B_{\frac{3}{2},\frac{1}{2};\frac{1}{2}},
\end{align} where $A_{j;m_{j}}$ are the diagonal elements of the
spectral-function matrix in the $j$ basis, while $B_{j,j';m_j}$ is an
off-diagonal element ($j=j^\prime\pm1/2, m_j=m_{j'}=1/2$); the latter turns out
to be small at low energy. \\

This provides a full description of the $t_{2g}$ manifold, which can recover a
pseudo-spin 1/2 system as a special case, but covers a broader class of
potential models. Such models combined with photoemission have previously been
used to gain significant understanding of Sr$_{2}$RuO$_{4}$, which shares a
similar amount of complexity associated with its low energy structure
\cite{Zhang2016, Kim2018}. The measured CPS ARPES intensity is proportional to
${\cal A}_{ \langle L_z S_z \rangle }$. Along $\Gamma \mathrm{Z}$, only
$A_{\frac{3}{2};\frac{3}{2}}$ and $A_{\frac{1}{2};\frac{1}{2}}$ contribute
sizably at very low energy, and thus determine the sign of
${\cal A}_{ \langle L_z S_z \rangle }$. Since the small $k_{z}$ dispersion is hard to resolve in our DMFT calculations,
for a quantitative analysis in \cref{fig:dmftspectral} we compare the DMFT
results to CPS-ARPES spectra integrated over the $k_{z}$ axis, finding excellent
agreement. Barring the precise energies where the sign of $\expval{L_z S_z}$
changes, all the positive and negative regions -- including the unexpected
negative peak around $E = -0.5$ eV -- are reproduced (and in fact the
quantitative agreement for the oscillating character of $\expval{L_z S_z}$ can
be observed, not only for the lowest energy states, but on the full 4 eV energy
window probed in the experiment). All the oscillations
are present in both panels, which implies that DMFT gives an accurate
representation of the band structure of Sr$_{2}$IrO$_{4}$. \\

\begin{figure}[b]
 \centering
\includegraphics[width=85mm]{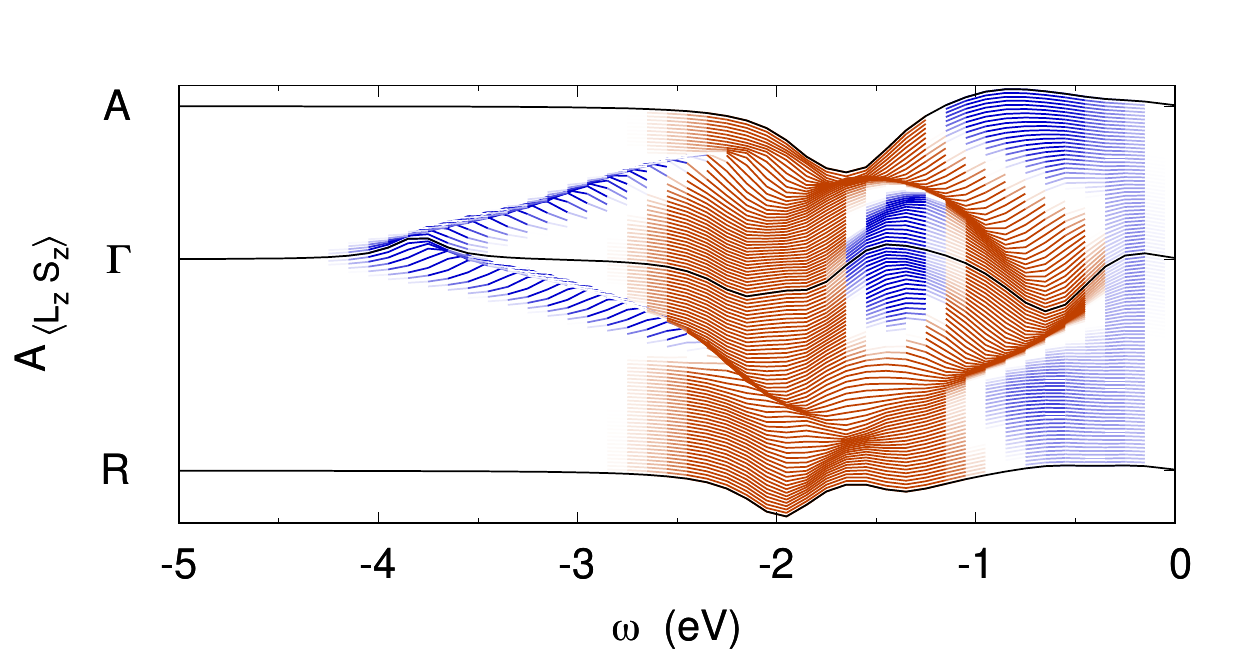}
\caption{\label{fig:dmftspectral_kdep} Evolution of $\expval{L_{z}S_{z}}$ along
  the directions $\Gamma - A$ and $\Gamma - R$, where
  $A=(\pi/a,\pi/a,\pi/c)$ and $R=(0,\pi/a,\pi/c)$. }
\end{figure}

It is worth pointing out that, although the exact energy where the sign changes
deviates, the presence of these oscillations itself is independent of the
precise values of the Coulomb parameters $U$ and $J$, or often adopted
approximations of the Coulomb vertex, provided that they yield an insulating
state. This further supports our conclusion that taking into account the
multi-band nature of the system is the critical starting point. Our DMFT
calculations give us direct access to the projections onto the $j_{1/2}$ and
$j_{3/2}$ states (shown as bare lines in \cref{fig:dmftspectral}), allowing us
to make more substantiated comments about our earlier claims: the spectral
weight in the low energy states arises approximately for 74\% from $j_{3/2}$
states, determined as the ratio of spectral weights integrated from the sign
change in $\expval{L_z S_z}$ at $E = -1.64$ eV all the way up to $E = 0$ eV.\\

The results obtained along ${\Gamma}\mathrm{A}$ and $\Gamma\mathrm{R}$ are shown
in \cref{fig:dmftspectral_kdep}; away from the $\Gamma\mathrm{Z}$ direction.
Since these results are presented away from ``normal emission'', to compare
these directly to experiments (i.e \cref{fig:k_dep_CPSA}), a more advanced
treatment of the photoemission dipole matrix elements is required. From these
plots, we can see that towards the zone boundaries, the overall magnitude of
$\expval{L_{z}S_{z}}$ decreases, as we observe in \cref{fig:k_dep_CPSA}, and has
also been suggested in~\cite{Louat2018} on the basis of density functional
theory calculations. \\

\section{Conclusions}

With our experiments and accompanying DMFT analysis we have thus demonstrated that a
pseudo-spin 1/2 model is insufficient to give a satisfying description of the
system at hand. Rather, one needs to rely on a modeling in terms of at least
the full $t_{2g}$ states and electron-electron interactions. While a description in
terms of $j_{1/2}$ orbitals was instrumental in developing our initial
understanding of Sr$_{2}$IrO$_{4}$ and of why spin orbit coupling gives rise to
an insulating ground state \cite{Kim2008, Zwartsenberg2020}, it is clear that this
model lacks the descriptive power needed to make further reaching conclusions; especially
connections made to the superconducting cuprates should be reevaluated in this
light.
Finally, and most importantly, we have demonstrated that with a carefully
crafted combination of a sufficiently complete many-body computational framework
and state-of-the-art experimental approaches, we can make tangible
progress in understanding materials with closely intertwined energy scales such
as Sr$_{2}$IrO$_{4}$. \\

\section*{Acknowledgements}
This research was undertaken thanks in part to funding from the Max
Planck-UBC-UTokyo Centre for Quantum Materials and the Canada First Research
Excellence Fund, Quantum Materials and Future Technologies Program. This project
is also funded by the Canada Research Chairs Program (A.D.); Natural Sciences
and Engineering Research Council of Canada (NSERC), Canada Foundation for
Innovation (CFI); British Columbia Knowledge Development Fund (BCKDF); and CIFAR
Quantum Materials Program. E.R. acknowledges support from the Swiss National
Science Foundation (SNSF) grant no. P300P2 164649. G.Z. acknowledges financial
support by the National Natural Science Foundation of China under Grants No.
12074384 and No. 11774349. E.P. and G.Z. gratefully acknowledges computer grants
on the supercomputer JUWELS and JURECA at the J\"ulich Supercomputing Centre
(JSC). This research used resources of the Advanced Light Source, a U.S. DOE
Office of Science User Facility under contract no. DE-AC02-05CH11231. \\

\appendix
\section{Effective $j$ states}
The $j_{\mathrm{eff}}$ states arise from the similarities the $t_{2g}$ orbitals
share with the $p$ orbitals, in particular with relation to spin-orbit coupling.
We will construct the Hamiltonian, for which we first define the $t_{2g}$ basis:
\begin{equation}
b_{t_{2g}} =  \left\{d_{xy,\up},d_{xz,\up},d_{yz,\up},d_{xy,\dn},d_{xz,\dn},d_{yz,\dn}\right\},
\end{equation}
we get for the $H_{SOC}$:
\begin{equation}
H_{SOC,t_{2g}} =
\frac{\lambda}{2}\begin{pmatrix}
                   0  & 0 &  0 & 0 & -i & 1 \\
                   0  & 0 &  -i & i & 0 & 0 \\
                   0  & i &  0 & -1 & 0 & 0 \\
                   0  & -i &  -1 & 0 & 0 & 0 \\
                   i  & 0 &  0 & 0 & 0 & i \\
                   1  & 0 &  0 & 0 & -i & 0 \\
\end{pmatrix}.
\end{equation}
We then consider a transformation to a new basis of ``effective''
$m_{l} \in \{-1, 0, 1\}$ (reminiscent of $p$) orbitals, which we define as:
\begin{gather}
\ket{1_{\mathrm{eff}}} = \frac{1}{\sqrt{2}}\left(\ket{d_{yz}} + i\ket{d_{xz}}\right) = i\ket{\sharm{-1}{2}}, \\
\ket{0_{\mathrm{eff}}} = -\ket{d_{xy}} = -\frac{i}{\sqrt{2}}\left(\ket{\sharm{-2}{2}} - \ket{\sharm{2}{2}}\right),\\
\ket{-1_{\mathrm{eff}}} = \frac{1}{\sqrt{2}}\left(-\ket{d_{yz}} + i\ket{d_{xz}}\right) =  -i\ket{\sharm{1}{2}}.
\end{gather}
Within this basis, the $L^+$ and $L_z$ operators become:
\begin{gather}
L^{+}_{l_{\mathrm{eff}}} = B_{l_{\mathrm{eff}}}^{-1} L^+ B_{l_{\mathrm{eff}}}  = \sqrt{2}
\begin{pmatrix}
                    0 & -1 &  0 \\
                    0 & 0  & -1  \\
                    0  & 0 & 0\\
\end{pmatrix},\\
L_{z,{l_{\mathrm{eff}}}} = B_{l_{\mathrm{eff}}}^{-1} L_{z} B_{l_{\mathrm{eff}}}  =
\begin{pmatrix}
                    -1 & 0 &  0 \\
                    0 & 0  & 0  \\
                    0  & 0 & 1\\
\end{pmatrix}.
\end{gather}
These are identical to the respective matrices for the $\ell = 1$ orbitals,
except they are multiplied by $-1$, and thus behaving effectively as $\ell = -1$
states. If we use these $\ell = -1$ states to construct spin-orbit entangled
states known as the $j_{\mathrm{eff}}$ states, as was first proposed in
\cite{Kim2008}, we obtain as the spin-orbit coupling Hamiltonian:
\begin{equation}
H_{SOC,j_{\mathrm{eff}}} =
\begin{pmatrix}
                   1  & 0 &  0 & 0 & 0 & 0 \\
                   0  & 1 &  0 & 0 & 0 & 0 \\
                   0  & 0 &  -\frac{1}{2} & 0 & 0 & 0 \\
                   0  & 0 &  0 &   -\frac{1}{2}  & 0 & 0 \\
                   0  & 0 &  0 & 0 &   -\frac{1}{2}  & 0 \\
                   0  & 0 &  0 & 0 & 0 &   -\frac{1}{2}  \\
\end{pmatrix}.
\end{equation}
This is again equal to the spin-orbit coupling Hamiltonian for $\ell = 1$
states, up to a minus sign. To this end, the expectation values
$\expval{\vb{L}\cdot\vb{S}}$ are negative to what is expected from ``regular''
$j$-states. \\

\section{Further CPS background}

\subsection{Data taken away from normal emission}
So far, the only expectation value discussed is the one along the $z$ direction,
and the calculated expectation values are only valid at the $\Gamma$-point.
Despite this, the technique has been successfully applied away from $\Gamma$
\cite{Day2018}. The equations hold true as long as not too much weight comes
from final states with $m_l \neq 0$. Following the $k$-dependent spherical
harmonic in \cref{eqn:matrixelement}, these other components have a
dependence $\propto \left(1 - \cos^2\theta_k \right)$, where $\theta_k$ is the
angle of the photoemitted electron and the surface normal. In particular, if the
photon energy is large, this angle is relatively small.

\begin{figure}[t]
\centering
  \includegraphics[width = 55mm]{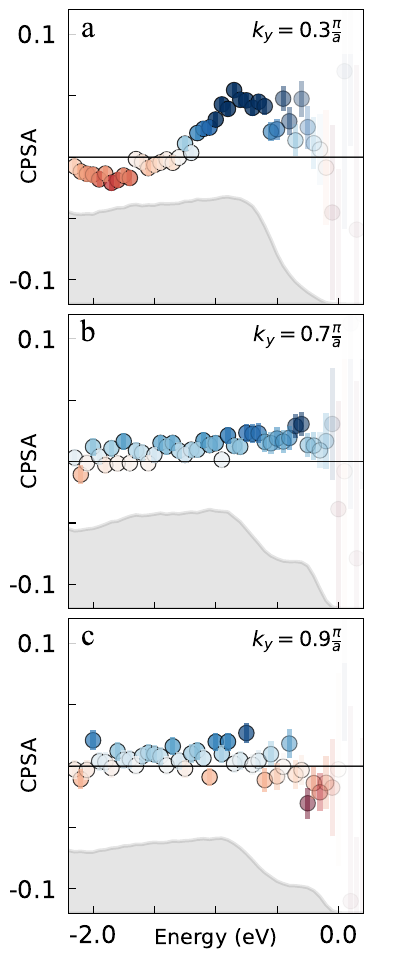}
  \caption{ (a-c) Measurement of the CPS ARPES signal through the Brillouin zone
    (colored markers) and spin-integrated signal (grey line). }
\label{fig:k_dep_CPSA}
\end{figure}

\subsection{Polarization of the incoming light}
The calculations presented up to this point have assumed that the incident light
is perfectly perpendicular to the surface. In the geometry of a realistic
ARPES experiment, the electron analyzer would be in the light path.
Therefore, the incidence angle of the light is usually approximately 45$^\circ$.
We will investigate here what effect of such an incidence angle is on the final
spectrum. \\
Taking the direction of the sample surface normal to be $\hat{\vb{z}}$, we can
write for the incoming light:
\[
\boldsymbol{\varepsilon}^{\oplus} = \varepsilon_0 \left( \frac{1}{\sqrt{4}}\left(\hat{\vb{x}} - \hat{\vb{z}} \right) + i\hat{\vb{y}}  \right).
\]
This can be converted into spherical harmonics, for which we can easily read off
the equivalent normal incidence light parameters:
\begin{figure}[b]
\centering
  \includegraphics[width = 55mm]{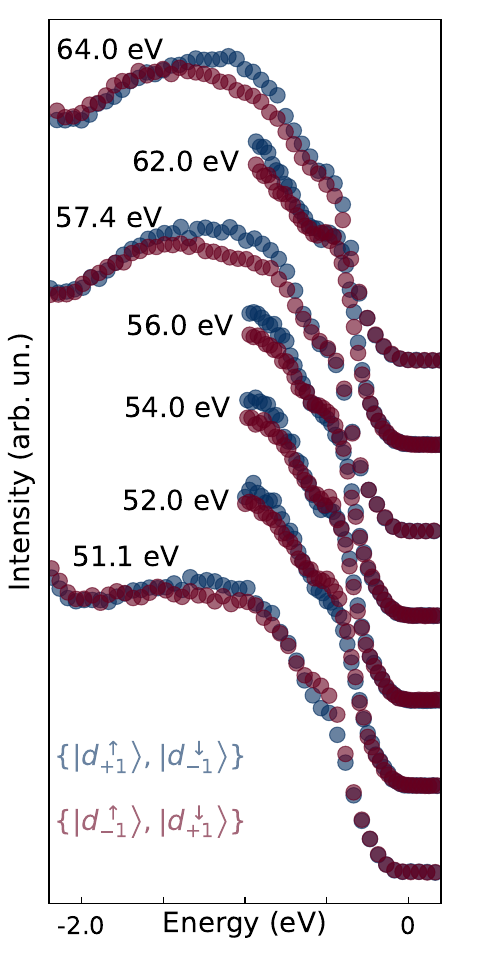}
  \caption{Photon energy dependent CPS ARPES measurements, plots of the parallel
    (blue) and antiparallel (red) signal.}
\label{fig:coan_kz}
\end{figure}
\[
  \boldsymbol{\varepsilon}^{\oplus} = \frac{1}{\sqrt{4}}\sharm{0}{1}  +  (\frac{1}{\sqrt{4}} +  \frac{1}{\sqrt{8}})\sharm{1}{1} +  (\frac{1}{\sqrt{4}} -  \frac{1}{\sqrt{8}})\sharm{-1}{1}.
\]
This deviates from the ideal case where we only make excitations with
$\sharm{1}{1}$. However, at $\Gamma$, there are no available final state
channels for $\sharm{0}{1}$ to scatter into. At finite emission angles $\theta$
this will generate a small unpolarized contribution that grows as
$1 - \cos^{2} \theta$. The $\sharm{-1}{1}$ term meanwhile creates excitations of
the opposite spin-orbital entanglement. Taking the squares of these
coefficients, we get $0.73$, and for $0.02$ for $\sharm{1}{1}$ and
$\sharm{-1}{1}$ respectively. This means that this configuration leads to an
opposite signal of just 3\% at normal emission, generating a net 6\% of
additional, unpolarized signal. This is far less than the approximate Sherman
function \cite{Bigi2017}, which is around 50\% for the (high-efficiency) VLEED
detectors we have used for our measurements. We can therefore safely ignore the
angle of the incoming light. \\

\section{Additional CPS data}
\subsection{In-plane $k$-dependent CPS ARPES data}
To identify how the spin-orbital entanglement changes throughout the Brillouin
zone, we present CPS ARPES measurements at various points along the
$(0,0) - (0,\pi)$ direction in \cref{fig:k_dep_CPSA}. Going away from the
$\Gamma$ point, the CPS ARPES signal rapidly diminishes until the signal
completely vanishes at the zone boundary. Similar observations are made if the
measurements are taken along the $(0,0) - (\pi,0)$ direction (not shown),
confirming the reliability of the measurement in this $C_4$ symmetric system.
Previous work has suggested that the coupling into $j_{\mathrm{eff}}$ states is
strongest at $\Gamma$, while hopping terms have a larger influence at the zone
boundaries \cite{Louat2018}, which is consistent with our observations. These
data support the interpretation that the spin-orbital entanglement varies
through $k$-space, and in fact reduces toward the zone boundary.

\subsection{Individual components of the CPS ARPES signal}
In order to better understand what regions in energy the features in the CPS
ARPES spectra arise, it is insightful to plot the parallel
(${\{\ket{d_{+1}^{\up}},\ket{d_{-1}^{\dn}}\}}$) and anti-parallel
(${\{\ket{d_{-1}^{\up}},\ket{d_{+1}^{\dn}}\}}$) components of the spectrum,
defined as:
\[
  I_{\{\ket{d_{+1}^{\up}},\ket{d_{-1}^{\dn}}\}} &=  \sqrt{I_{\up}^{\ominus}I_{\dn}^{\oplus}} \\
  I_{\{\ket{d_{-1}^{\up}},\ket{d_{+1}^{\dn}}\}} &=  \sqrt{I_{\dn}^{\ominus}I_{\up}^{\oplus}},
\]
which together form the CPS ARPES signal as defined in the main text. These signals
are plotted in \cref{fig:coan_kz} in blue ($I_{\{\ket{d_{+1}^{\up}},\ket{d_{-1}^{\dn}}\}}$) and red
($I_{\{\ket{d_{-1}^{\up}},\ket{d_{+1}^{\dn}}\}}$) for the same photon energies as presented in
Fig. 3 in the main text. From these spectra it is straightforward
to see that the sign-changing signal in $k_{z}$ arises from the state that
appears as a shoulder around $E = -0.5$ eV.
\vspace{2mm}

\clearpage

\end{document}